%% file: pub7099p.tex
\documentstyle[12pt]{article}
 
\renewcommand{\baselinestretch}{1.1}

\begin{document}

\font\elevenrm=cmr9 scaled\magstep1      \let\elrm=\elevenrm
  \font\elevensl=cmsl9 scaled\magstephalf  \let\elsl=\elevensl
    \font\elevenbf=cmbx9 scaled\magstep1     \let\elbf=\elevenbf
      \font\elevenit=cmti9 scaled\magstephalf   \let\elit=\elevenit
 
\font\twelvebsl=cmbxsl10 scaled\magstep 1
   \let\bsl=\twelvebsl
\def\mbf#1{\hbox{${\twelvebsl #1}$}}
\font\tenbsl=cmbxsl10
 
\font\eightrm=cmr8 scaled\magstep1
 
\font\ninerm=cmr9 scaled\magstep1
\catcode`\@=11 
\def\lsim{\mathrel{\mathpalette\@versim<}}
\def\gsim{\mathrel{\mathpalette\@versim>}}
\def\@versim#1#2{\vcenter{\offinterlineskip
        \ialign{$\m@th#1\hfil##\hfil$\crcr#2\crcr\sim\crcr } }}
%
\makeatletter
\def\@seccntformat#1{\csname the#1\endcsname.\hskip 1em}
\renewcommand\thesubsection{\Alph{subsection}}
\makeatother
 
  \def\str{\penalty-10000\hfilneg\ } 
   \def\nostr{\hfill\penalty-10000\ } 

\input mydefs

\def\Oa{O(\alpha_s)}
\def\Oaa{O(\alpha_s^2)}
 
\thispagestyle{empty}
 
\begin{flushright}
{\footnotesize\renewcommand{\baselinestretch}{.75}
  SLAC-PUB-7099\\
  May 1996\\
}
\end{flushright}
 
\vskip .5truecm
 
\begin{center}
{\Large \bf A STUDY OF THE ORIENTATION AND ENERGY PARTITION
OF THREE-JET EVENTS IN HADRONIC $Z^0$ DECAYS$^*$}
\end{center}
 
\vspace {1.0cm}

\vspace {1.0cm}
 
\begin{center}
{\bf The SLD Collaboration$^{\diamond}$}\\
{Stanford Linear Accelerator Center, Stanford University,}\\
{Stanford, CA\  94309}
\end{center}
 
\vfill
 
\begin{center}
Submitted to {\it Physical Review D}
\end{center}
 
\vfill
 
\noindent $^*${\footnotesize Work supported by
U.S. Department of Energy contracts:
  DE-FG02-91ER40676 (BU),
  DE-FG03-92ER40701 (CIT),
  DE-FG03-91ER40618 (UCSB),
  DE-FG03-92ER40689 (UCSC),
  DE-FG03-93ER40788 (CSU),
  DE-FG02-91ER40672 (Colorado),
  DE-FG02-91ER40677 (Illinois),
  DE-AC03-76SF00098 (LBL),
  DE-FG02-92ER40715 (Massachusetts),
  DE-AC02-76ER03069 (MIT),
  DE-FG06-85ER40224 (Oregon),
  DE-AC03-76SF00515 (SLAC),
  DE-FG05-91ER40627 (Tennessee),
  DE-AC02-76ER00881 (Wisconsin),
  DE-FG02-92ER40704 (Yale);
  U.S. National Science Foundation grants:
  PHY-91-13428 (UCSC),
  PHY-89-21320 (Columbia),
  PHY-92-04239 (Cincinnati),
  PHY-88-17930 (Rutgers),
  PHY-88-19316 (Vanderbilt),
  PHY-92-03212 (Washington);
  the UK Science and Engineering Research Council
  (Brunel and RAL);
  the Istituto Nazionale di Fisica Nucleare of Italy
  (Bologna, Ferrara, Frascati, Pisa, Padova, Perugia);
  and the Japan-US Cooperative Research Project on High Energy Physics
  (Nagoya, Tohoku).}
 
\eject

%
%
 
\vspace*{-.35in}
%
%
%
  \def\iADEL{$^{(1)}$}
  \def\iBOL{$^{(2)}$}
  \def\iBU{$^{(3)}$}
  \def\iBRUN{$^{(4)}$}
  \def\iCIT{$^{(5)}$}
  \def\iUCSB{$^{(6)}$}
  \def\iUCSC{$^{(7)}$}
  \def\iCIN{$^{(8)}$}
  \def\iCSU{$^{(9)}$}
  \def\iCOLO{$^{(10)}$}
  \def\iCOL{$^{(11)}$}
  \def\iFER{$^{(12)}$}
  \def\iFRA{$^{(13)}$}
  \def\iILL{$^{(14)}$}
  \def\iLBL{$^{(15)}$}
  \def\iMIT{$^{(16)}$}
  \def\iMASS{$^{(17)}$}
  \def\iMISS{$^{(18)}$}
  \def\iMOSC{$^{(19)}$}
  \def\iNAG{$^{(20)}$}
  \def\iOREG{$^{(21)}$}
  \def\iPAD{$^{(22)}$}
  \def\iPERU{$^{(23)}$}
  \def\iPISA{$^{(24)}$}
  \def\iRUT{$^{(25)}$}
  \def\iRAL{$^{(26)}$}
  \def\iSOGANG{$^{(27)}$}
  \def\iSLAC{$^{(28)}$}
  \def\iTENN{$^{(29)}$}
  \def\iTOH{$^{(30)}$}
  \def\iVAND{$^{(31)}$}
  \def\iWASH{$^{(32)}$}
  \def\iWISC{$^{(33)}$}
  \def\iYALE{$^{(34)}$}
  \def\dead{$^{\dag}$}
  \def\andgen{$^{(a)}$}
  \def\andper{$^{(b)}$}
%
\begin{center}
 
\small
 
\baselineskip=.8\baselineskip   
 
\mbox{$\diamond$ K. Abe                 \unskip,\iNAG}
\mbox{K. Abe                 \unskip,\iTOH}
\mbox{I. Abt                 \unskip,\iILL}
\mbox{T. Akagi               \unskip,\iSLAC}
\mbox{N.J. Allen             \unskip,\iBRUN}
\mbox{W.W. Ash               \unskip,\iSLAC$^\dagger$}
\mbox{D. Aston               \unskip,\iSLAC}
\mbox{K.G. Baird             \unskip,\iRUT}
\mbox{C. Baltay              \unskip,\iYALE}
\mbox{H.R. Band              \unskip,\iWISC
% ...
}
\end{center}
 
\vfill
\eject
 
\begin{center}
\small
\it
%
%
%
  \baselineskip=.75\baselineskip   
  \iADEL
     Adelphi University,
     Garden City, New York 11530 \break
  \iBOL
     INFN Sezione di Bologna,
     I-40126 Bologna, Italy \break
  \iBU
     Boston University,
     Boston, Massachusetts 02215 \break
  \iBRUN
     Brunel University,
     Uxbridge, Middlesex UB8 3PH, United Kingdom \break
  \iCIT
     California Institute of Technology,
     Pasadena, California 91125 \break
  \iUCSB
     University of California at Santa Barbara,
     Santa Barbara, California 93106 \break
  \iUCSC
     University of California at Santa Cruz,
     Santa Cruz, California 95064 \break
  \iCIN
     University of Cincinnati,
     Cincinnati, Ohio 45221 \break
  \iCSU
     Colorado State University,
     Fort Collins, Colorado 80523 \break
  \iCOLO
     University of Colorado,
     Boulder, Colorado 80309 \break
  \iCOL
     Columbia University,
     New York, New York 10027 \break
  \iFER
     INFN Sezione di Ferrara and Universit\`a di Ferrara,
     I-44100 Ferrara, Italy \break
  \iFRA
     INFN  Lab. Nazionali di Frascati,
     I-00044 Frascati, Italy \break
  \iILL
     University of Illinois,
     Urbana, Illinois 61801 \break
  \iLBL
     Lawrence Berkeley Laboratory, University of California,
     Berkeley, California 94720 \break
  \iMIT
     Massachusetts Institute of Technology,
     Cambridge, Massachusetts 02139 \break
  \iMASS
     University of Massachusetts,
     Amherst, Massachusetts 01003 \break
  \iMISS
     University of Mississippi,
     University, Mississippi  38677 \break
  \iMOSC
     Moscow State University,
     Institute of Nuclear Physics,
     119899 Moscow,
     Russia    \break
  \iNAG
     Nagoya University,
     Chikusa-ku, Nagoya 464 Japan  \break
  \iOREG
     University of Oregon,
     Eugene, Oregon 97403 \break
  \iPAD
     INFN Sezione di Padova and Universit\`a di Padova,
     I-35100 Padova, Italy \break
  \iPERU
     INFN Sezione di Perugia and Universit\`a di Perugia,
     I-06100 Perugia, Italy \break
  \iPISA
     INFN Sezione di Pisa and Universit\`a di Pisa,
     I-56100 Pisa, Italy \break
  \iRUT
     Rutgers University,
     Piscataway, New Jersey 08855 \break
  \iRAL
     Rutherford Appleton Laboratory,
     Chilton, Didcot, Oxon OX11 0QX United Kingdom \break
  \iSOGANG
     Sogang University,
     Seoul, Korea \break
  \iSLAC
     Stanford Linear Accelerator Center, Stanford University,
     Stanford, California 94309 \break
  \iTENN
     University of Tennessee,
     Knoxville, Tennessee 37996 \break
  \iTOH
     Tohoku University,
     Sendai 980 Japan \break
  \iVAND
     Vanderbilt University,
     Nashville, Tennessee 37235 \break
  \iWASH
     University of Washington,
     Seattle, Washington 98195 \break
  \iWISC
     University of Wisconsin,
     Madison, Wisconsin 53706 \break
  \iYALE
     Yale University,
     New Haven, Connecticut 06511 \break
  \dead
     Deceased \break
  \andgen
     Also at the Universit\`a di Genova \break
  \andper
     Also at the Universit\`a di Perugia \break
\rm
%
 
\end{center}

\eject

\centerline{\bf ABSTRACT}
 
\vskip .5truecm
 
\noindent
{\small
We have measured the distributions of the
jet energies in \ep \ra \qqg events, and of the
three orientation angles of the event plane,
using hadronic $Z^0$ decays collected in the SLD experiment at SLAC.
We find that the data are well described
by perturbative QCD incorporating
vector gluons. We have also compared our data with models of scalar
and tensor gluon production, and discuss
limits on the relative contributions
of these particles to three-jet production in \epa.
}
 
\vskip 1.5truecm

\section{Introduction}

The observation of \epa into final states containing three hadronic
jets \cite{gluon}, and their interpretation
in terms of the process \ep \ra \qqg, provided the first direct
evidence for the existence of the gluon, the gauge boson
of the theory of strong interactions, Quantum Chromodynamics (QCD)
\cite{qcd}. Following these initial observations
studies of the partition of energy among the three jets
were performed at the PETRA and PEP storage rings.
Comparison of the data with leading-order
QCD predictions, and with a model incorporating the radiation of
spin-0 (scalar) gluons, provided qualitative evidence \cite{PETRA}
for the spin-1 (vector) nature of the gluon, which is a fundamental
element of QCD. Similar studies have since been
performed at LEP \cite{LTHREE,OPAL}.
 
An additional interesting observable in three-jet events is the
orientation of the event plane w.r.t. the beam direction, which can be
described by three Euler angles. These angular distributions were
studied first by TASSO \cite{TASSO}, and more recently by
L3 \cite{LTHREE} and DELPHI \cite{DELPHI}.
Again, the data were compared with the predictions of perturbative
QCD and a scalar gluon model, but the Euler angles
are less sensitive than the jet energy distributions
to the differences between the two cases \cite{LTHREE}.
 
Here we present measurements of the jet energy
and event plane orientation
angle distributions from hadronic decays of $Z^0$
bosons produced by e$^+$e$^-$ annihilations at
the SLAC Linear Collider (SLC) and recorded in the SLC Large Detector
(SLD). We used particle energy deposits measured in
the SLD Liquid Argon Calorimeter, which covers 98\% of
the solid angle, for jet reconstruction.
We compare our measured distributions with the predictions of
perturbative QCD and a scalar gluon model. In addition, we make
the first comparison \cite{hwang}
with a model which comprises spin-2 (tensor)
gluons, and discuss limits on the
possible relative contributions of scalar and tensor gluons to
three-jet production in \epa.
 
In Section 2 the observables are defined, and
the predictions of perturbative QCD and of the scalar and tensor gluon
models are discussed.
We describe the detector, the event trigger, and the selection criteria
applied to the data, in Section 3. The three-jet analysis is
described in Section 4, and a summary and conclusions are presented
in Section~5.
 
\section{Observables and Theoretical Predictions}
 
\subsection{Scaled Jet Energy Distributions}
 
Ordering the three jets in \ep \ra \qqg $\;$ according to their energies,
$E_1>E_2>E_3$, and normalising by the c.m. energy $\sqrt{s}$, we
obtain the scaled jet energies:
\begin{eqnarray}
x_i\quad=\quad {2 E_i\over\sqrt{s}}\quad\quad\quad(i=1,2,3),
\end{eqnarray}
where $ x_1 +  x_2 +  x_3 = 2$.
Making a Lorentz boost of the event into the rest frame of
jets 2 and 3 the Ellis-Karliner angle $\theta_{EK}$ is defined \cite{ek}
to be the angle between jets 1 and 2 in
this frame. For massless partons at tree-level:
\begin{eqnarray}
cos\theta_{EK}={{ x_2- x_3}\over{ x_1}}.
\end{eqnarray}
The inclusive differential cross section can be calculated to
$\Oa$ in perturbative QCD incorporating spin-1
(vector) gluons and assuming massless partons \cite{kramer}:
\begin{eqnarray}
 {1\over\sigma}{d^2\sigma^V\over d x_1d x_2} \propto
{{ x_1^3+ x_2^3+(2- x_1- x_2)^3}\over{(1- x_1)(1- x_2)( x_1+ x_2-1)}}.
\end{eqnarray}
 
One can also consider alternative `toy' models of
strong interactions. For a model incorporating spin-0
(scalar) gluons one obtains at leading order at the \z0 resonance
\cite{scalar}:
\begin{eqnarray}
 {1\over\sigma}{d^2\sigma^S\over d x_1d x_2} \propto
\biggl[{{ x_1^2(1- x_1)+ x_2^2(1- x_2)+(2- x_1- x_2)^2( x_1+ x_2-1)}
\over  {(1- x_1)(1- x_2)( x_1+ x_2-1)}}-R \biggr]
\end{eqnarray}
where
\begin{eqnarray}
 {R}={10\;{\Sigma_j a_j^2}\over{\Sigma_j (v_j^2+a_j^2)}}
\end{eqnarray}
and $a_j$ and $v_j$ are the axial and vector couplings, respectively,
of quark flavor $j$ to the \z0.
For a model of strong interactions incorporating spin-2
(tensor) gluons (see Appendix) one obtains at leading order:
\begin{eqnarray}
 {1\over\sigma}{d^2\sigma^T\over d x_1d x_2} \propto
{{(x_1+x_2-1)^3 + (1-x_1)^3 + (1-x_2)^3}
\over{(1- x_1)(1- x_2)( x_1+ x_2-1)}}.
\end{eqnarray}
 
Singly-differential cross sections for $x_1$, $x_2$, $x_3$
or cos$\theta_{EK}$ were obtained by numerical
integrations of Eqs.~3, 4 and 5.
These cross sections are shown in Fig.~1; the shapes are different for
the vector, scalar and tensor gluon cases.
 
It is well known that vector particles coupling to quarks
in either Abelian or
non-Abelian theories allow consistent and renormalizable
calculations to all orders in perturbation theory. However,
the scalar and tensor gluon models have limited
applicability beyond leading order.
In the scalar model no symmetry, such as gauge invariance, exists to
prevent the gluons from acquiring mass. In the tensor case the
model is non-renormalizable (see Appendix), so
that higher order predictions are not physically meaningful. Given
these difficulties we limit ourselves to the leading-order expressions
for 3-jet event production in these two cases. In the vector case we
do consider the influence of higher-order corrections to the
leading-order predictions. We also assume that the transformation of
the partons in 3-jet events into the observed hadrons is independent
of the gluon spin.
 
\subsection{Event Plane Orientation}
 
The orientation of the three-jet event plane can be described by the
angles $\theta$, $\theta_N$ and $\chi$ illustrated in Fig.~2.
When no explicit quark, antiquark or gluon jet identification is made,
$\theta$ is the polar angle of the most energetic
jet w.r.t. the electron
beam direction, $\theta_N$ is the polar angle of the normal to the
event plane w.r.t. the electron beam direction, and $\chi$ is the
angle between the event plane and the plane containing the electron
beam and the most energetic jet.
The distributions of these angles may be written \cite{scalar}:
\begin{eqnarray}
{d\sigma\over d{\rm cos}\theta}\quad\propto\quad 1\;+\;\alpha(T)
{\rm cos}^2\theta
\end{eqnarray}
\begin{eqnarray}
{d\sigma\over d{\rm cos}\theta_N}\quad\propto\quad 1\;+\;\alpha_N(T)
{\rm cos}^2\theta_N
\end{eqnarray}
\begin{eqnarray}
{d\sigma\over d\chi}\quad\propto\quad 1\;+\;\beta(T)
{\rm cos}2\chi
\end{eqnarray}
where $T$ is the thrust value \cite{thrust}
of the event. The coefficients $\alpha(T)$,
$\alpha_N(T)$ and $\beta(T)$ depend on the gluon spin; they are
shown in Fig.~16 for leading-order calculations incorporating vector,
scalar and tensor gluons.
In perturbative QCD \oalpsq corrections to the leading-order result
have been calculated and are small \cite{ksb}.
 
In \z0 decay events produced with longitudinally-polarized
electrons an additional term $\beta_N S_Z$cos$\theta_N$, representing
a correlation between the event-plane orientation and the \z0 spin
direction, should be added to eq.~(8). For Standard Model processes
the correlation parameter
$\beta_N$ is expected \cite{Lance} to be of order $10^{-5}$, which is
well below our current experimental sensitivity \cite{Todd}.
In this analysis we have ignored information on the helicity of the
electron beam and are hence insensitive to a term in eq.~(8)
linear in cos$\theta_N$.
 
\section{Apparatus and Hadronic Event Selection}

The e$^+$e$^-$ annihilation events produced at the $Z^0$ resonance
by the SLC in the 1993 run were recorded using the SLD.
A general description of the SLD can be found elsewhere \cite{sld}.
The analysis presented here used particle energy deposits measured
in the Liquid Argon Calorimeter (LAC) \cite{lac},
which contains both electromagnetic and hadronic sections,
and in the Warm Iron Calorimeter \cite{wic}.
The trigger for hadronic events required
a total LAC electromagnetic energy greater than 12 GeV.
 
Clusters were
formed from the localized energy depositions in the LAC; energy
depositions consistent with
background muons produced upstream in the accelerator
were identified and removed \cite{johny}.
The measured cluster energies were then corrected \cite{hwang} for the
response of the LAC, which
varies with polar angle $\theta$ due to the material of the
inner detector components as well as the thinner calorimeter
coverage at the endcap-barrel interface, using a detailed Monte Carlo
simulation of the detector.
We first verified that the measured energy of clusters in
each polar-angle bin, integrated over all selected clusters
in all selected hadronic events, was well described by the simulation.
Next, the ratio of simulated cluster energy to generated
particle energy was calculated for each cluster. This ratio was
averaged over all clusters in each polar-angle bin to yield the
response function $r(\theta)$.
Finally, the measured energy of each cluster in the data was weighted
by $1/r(\theta)$. The normalised r.m.s. deviation
of the distribution of the total
cluster energy in hadronic events was 21\% before, and 16.5\% after,
application of this procedure \cite{hwang}.
 
Corrected
clusters were then required to have a non-zero electromagnetic energy
component and a total energy $E_{cl}$ of at least 100 MeV.
For each event the total cluster energy $E_{tot}$, energy imbalance
$\Sigma|\vec{E_{cl}}|/E_{tot}$, and thrust axis polar angle $\theta_T$
\cite{thrust} were calculated from the selected corrected clusters.
Events with $|{\rm cos}\theta_T|\leq0.8$
($|{\rm cos}\theta_T|\geq0.8$)
were then required to contain at least 8 (11) such clusters,
to have $E_{tot}$ $>$ 15 GeV, and to have
$\Sigma|\vec{E_{cl}}|/E_{tot}$ $<$ 0.6.
From our 1993 data sample approximately 51,000
events passed these cuts. The efficiency  for selecting hadronic events
was estimated to be $92\pm2\%$, with an estimated
background in the selected sample of $0.4\pm 0.2\%$ \cite{alr},
dominated by $Z^0 \rightarrow \tau^+ \tau^-$
and $Z^0$ $\rightarrow$ \ep events.
 
 
\section{Data Analysis}
 
Jets were reconstructed from selected LAC clusters in selected
hadronic events. The JADE jet-finding algorithm \cite{JADE}
was used, with a scaled invariant mass cutoff value $y_c$ = 0.02, to
identify a sample of 22,114 3-jet events. This $y_c$ value
maximises the rate of events classified as 3-jet final states;
other values of $y_c$ were also considered and found not to
affect the conclusions of this study.
A non-zero sum of the three
jet momenta can be induced in the selected events
by particle losses due to the acceptance and inefficiency of the
detector, and by jet energy resolution effects. This was corrected by
rescaling the measured jet momenta $\vec{P_i}$
($i$ = 1,2,3) according to:
\begin{eqnarray}
{P_i^{j}}^{\prime}\quad=\quad P_i^j - R^j |P_i^j|
\end{eqnarray}
where $P_i^j$ is the $j$-th momentum component of jet $i$, $j$ =
$x$, $y$, $z$, and
\begin{eqnarray}
R^j\quad=\quad {\Sigma_{i=1}^3 P_i^j \over \Sigma_{i=1}^3 |P_i^j|}.
\end{eqnarray}
The jet energy components were then rescaled according to:
\begin{eqnarray}
{E_i}^{\prime}\quad=\quad {|\vec{P_i}^{\prime}|\over{|\vec{P_i}|}}\;E_i
\end{eqnarray}
This procedure resulted in a slight improvement in
the experimental resolution of the scaled jet energies $x_i$
\cite{hwang}.
 
\subsection{Scaled Jet Energy Distributions}
 
The measured distributions of the three scaled jet energies $x_1$,
$x_2$, $x_3$, and the Ellis-Karliner angle $\theta_{EK}$, are
shown in Fig. 3. Also shown in Fig.~3 are
the predictions of the
HERWIG 5.7 \cite{HERWIG} Monte Carlo program for the simulation of
hadronic decays of $Z^0$ bosons, combined with a simulation of the SLD
and the same selection and analysis cuts as applied to the real data.
The simulation describes the data well.
 
For each observable $X$,
the experimental distribution $D^{data}_{SLD}(X)$ was then
corrected for the effects of selection cuts,
detector acceptance, efficiency,
resolution, particle decays and interactions
within the detector, and for initial state photon radiation, using
bin-by-bin correction factors $C_D(X)$:
\begin{eqnarray}
C_D(X)_m =  \frac{D^{MC}_{hadron}(X)_m}{D^{MC}_{SLD}(X)_m},
\label{cd}
\end{eqnarray}
where: $m$ is the bin index;
$D^{MC}_{SLD}(X)_m$ is the content of bin $m$ of the distribution
obtained from reconstructed clusters in Monte Carlo events
after simulation of the detector; and
$D^{MC}_{hadron}(X)_i$ is that from all generated particles with
lifetimes greater than $3 \times 10^{-10}$~s in Monte Carlo events
with no SLD simulation and no initial state radiation.
The bin widths were chosen from the estimated experimental resolution
so as to minimize bin-to-bin migration effects.
The $C_D(X)$ were calculated from events generated with HERWIG 5.7
using default parameter values \cite{HERWIG}.
The {\it hadron level} distributions are then given by
\begin{eqnarray}
D^{data}_{hadron}(X)_m = C_D(X)_m \cdot D^{data}_{SLD}(X)_m.
\end{eqnarray}
 
Experimental systematic errors arising from uncertainties in
modelling the detector were estimated by varying the
event selection criteria over wide ranges, and by varying the cluster
energy response corrections in the detector simulation \cite{hwang}.
In each case the correction factors $C_D(X)$, and hence the corrected
data distributions $D^{data}_{hadron}(X)$, were rederived.
The correction
factors $C_D(X)$ are shown in Figs. 4(b)--7(b);
the errors comprise the sum in quadrature of the
statistical component from the finite size of the Monte Carlo event
sample, and the systematic uncertainty. It can be seen that the
$C_D(X)$ are close to unity and slowly-varying, except near the
boundaries of phase-space. The hadron level
data are listed in Tables I--IV, together with statistical
and systematic errors; the central values represent the data corrected
by the central values of the correction factors.
 
Before they can be compared with parton-level predictions the data must
be corrected for the effects of hadronization.
In the absence of a complete theoretical calculation,
the phenomenological models implemented in JETSET 7.4 \cite{JETSET}
and HERWIG 5.7 represent our best description of the
hadronization process, and are not based upon a particular choice of
the gluon spin. These models have been compared
extensively with, and tuned to, \ep \ra hadrons data at the $Z^0$
resonance \cite{ztune}, as well as data at $W$ $\sim$ 35 GeV from
the PETRA and PEP storage rings \cite{peptune}. We find that
they provide a
good description of our data in terms of the observables presented
here (Fig.~3) and other hadronic event shape observables
\cite{sldevsh}, and hence employ them to calculate hadronization
correction factors. The HERWIG parameters were left at their default
values. Several of the JETSET parameters were set to values
determined from our own optimisation to hadronic $Z^0$ data; these are
given in Table V.
 
The hadronization
correction procedure is similar to that described above for
the detector effects. Bin-by-bin correction factors
\begin{eqnarray}
C_H(X)_m =  \frac{D^{MC}_{parton}(X)_m}{D^{MC}_{hadron}(X)_m},
\end{eqnarray}
where $D^{MC}_{parton}(X)_m$ is the
content of bin $m$ of the distribution obtained from Monte Carlo events
generated at the parton level, were calculated and applied to
the hadron level data distributions $D^{data}_{hadron}(X)_m$ to
obtain the {\it parton level} corrected data:
\begin{eqnarray}
D^{data}_{parton}(X)_m = C_H(X)_m \cdot D^{data}_{hadron}(X)_m.
\end{eqnarray}
For each bin
the average of the JETSET-- and HERWIG--derived values was used as the
central value of the correction factor, and the difference between this
value and the extrema was assigned as a symmetric hadronization
uncertainty. The correction
factors $C_H(X)$ are shown in Figs. 4(c)--7(c);
the errors comprise the sum in quadrature of the
statistical component from the finite size of the Monte Carlo event
sample, and the systematic uncertainty.
It can be seen that the $C_H(X)$ are within 10\% of unity and are
slowly-varying, except near the boundaries of phase space.
The fully-corrected data are shown in Figs.~4(a)--7(a);
the data points
correspond to the central values of the correction factors,
and the errors shown comprise the
statistical and total systematic components added in quadrature.
These results are in agreement with an analysis of our 1992 data
sample using charged tracks for jet reconstruction \cite{fan}.
 
We first compare the data with QCD predictions from
$\Oa$ and $\Oaa$ perturbation theory, and from parton shower (PS)
models. For this purpose we used the JETSET 7.4
$\Oa$ matrix element, $\Oaa$ matrix element, and PS
options, and the HERWIG 5.7 PS,
and generated events at the parton level.
In each case all parameters were left at their
default values \cite{HERWIG,JETSET}, with the exception of the
JETSET parton shower parameters listed in Table V. The QCD scale
parameter values used were $\Lambda$ = 1.0 GeV ($\Oa$), 0.25 GeV
(\oalpsq), 0.26 GeV (JETSET PS) and 0.18 GeV (HERWIG PS).
The shapes of the $x_1$, $x_2$, $x_3$ and cos$\theta_{EK}$
distributions do not depend on $\Lambda$ at $\Oa$, and only weakly so
at higher order. The resulting
predictions for $x_1$, $x_2$, $x_3$ and cos$\theta_{EK}$ are
shown in Figs. 4(a) -- 7(a). These results represent Monte Carlo
integrations of the respective QCD formulae and are hence
equivalent to analytic or numerical QCD results based on the same
formulae; in the $\Oa$ case we have checked explicitly that JETSET
reproduces the numerical results of the analytic calculation
described in Section 2.
 
The $\Oa$ calculation describes the data reasonably well, although
small discrepancies in the details of the shapes of the distributions
are apparent and the $\chi^2$ for the comparison between data and
MC is poor (Table VI).
The $\Oaa$ calculation describes the $x_1$, $x_2$ and $x_3$ data
distributions better, but the description of the cos$\theta_{EK}$
distribution is slightly worse; this is difficult to
see directly in Figs.~4(a)--7(a), but is evident from the $\chi^2$
values for the data--MC comparisons (Table VI).
Both parton shower calculations describe the
data better than either the $\Oa$ or $\Oaa$ calculations
and yield relatively good $\chi^2$ values (Table VI).
This improvement in the quality of description of the data
between the $\Oa$ and parton shower calculations
can be interpreted as an indication of the
contribution of multiple soft gluon emission to the fine details of the
shapes of the distributions.
In fact for all calculations the largest discrepancies,
at the level of at most 10\%, arise in the regions $x_1$ $>$ 0.98,
$x_2$ $>$ 0.93, $x_3$ $<$ 0.09 and cos$\theta_{EK}$ $>$
0.9, near the boundaries of phase space where soft and collinear
divergences are expected to be large and to require resummation in
QCD perturbation theory \cite{resum}; such resummation has not been
performed for the observables considered here.
 
For each observable we chose a range such that the
detector and hadronization correction factors are close to unity,
$0.8<C_D(X),C_H(X)<1.2$, have small uncertainty,
$\Delta C_D(X),\Delta C_H(X)<0.2$, and are
slowly-varying (see Figs.~4--7).
The ranges are: $0.688 < x_1 <0.976$, $x_2 <0.93$, $x_3>0.09$ and
cos$\theta_{EK}<0.9$; they exclude the phase-space boundary regions.
Within these ranges the comparison between data and calculations
yields significantly improved $\chi^2$ values
(values in parentheses in Table VI); the $\Oaa$ calculation has
acceptable $\chi^2$ values and those for both parton shower models are
typically slightly better. These results
support the notion that QCD, incorporating vector gluons, is the
correct theory of strong interactions.
 
We now consider alternative models of strong interactions,
incorporating scalar and tensor gluons, discussed in Section 2.
Since these model calculations are at leading order in perturbation
theory we also consider
first the vector gluon (QCD) case at the same order.
The data within the selected ranges are shown in Fig.~8;
from comparison with the raw data (Fig.~3) it is apparent
that the shapes of the distributions are barely affected by the
detector and hadronization corrections. The leading-order scalar, vector
and tensor gluon predictions, normalised to the
data within the same ranges, are also shown in Fig.~8.
The vector calculation clearly
provides the best description of the data; neither the scalar nor
tensor cases predicts the correct shape for any of the observables.
The $\chi^2$ values for the comparisons are given in Table VII.
This represents the first comparison
of a tensor gluon calculation with experimental data.
 
It is interesting to consider whether the data allow an admixture of
contributions from the different gluon spin hypotheses.
For this purpose
we performed simultaneous fits to a linear combination of the
vector (V) + scalar (S) + tensor (T) predictions,
allowing the relative normalisations to vary according to:
\begin{eqnarray}
(1-a-b)\, V \quad + \quad a\, S\quad + \quad b \, T
\end{eqnarray}
where $a$ and $b$ are free parameters determined from the fit.
For the vector contribution we used in turn the \oalp, \oalpsq, JETSET
PS and HERWIG PS calculations. In all cases the fit to the
distribution of each observable yielded a slightly lower $\chi^2$
value than the vector-only fit.
We found that the allowed contributions of scalar and tensor
gluons depend upon the order of the vector calculation used, as well
as on the observable.
The largest allowed scalar contribution was $a$ = 0.11 from the
fit to cos$\theta_{EK}$ using the \oalpsq calculation.
The largest allowed tensor contribution was $b$ = 0.31 from the
fit to $x_1$ using the $\Oa$ calculation.
The smallest allowed contributions were
$a$ and $b$ $<$ 0.001 from the fit to $x_1$ using the HERWIG PS.
 
Any pair of the observables $x_1$, $x_2$, $x_3$ and cos$\theta_{EK}$
may be taken to be independent variables, subject to the overall
constraint $x_1+x_2+x_3=2$. Therefore, in order to utilise more
information, we also performed fits of Eq.~17 simultaneously to the
$x_2$ and $x_3$ distributions. We found
the relative S, V and T contributions and the
$\chi^2/d.o.f.$ values to be comparable with those from
the fits to $x_2$ alone.
 
 
\subsection{Event Plane Orientation}
 
We now consider the three Euler angles that describe the orientation
of the event plane: $\theta$, $\theta_N$, and $\chi$ (Fig.~2).
The analysis procedure is similar to that described in the previous
section. The measured distributions of these angles are
shown in Fig. 9, together with the predictions of HERWIG 5.7,
combined with a simulation of the SLD
and the same selection and analysis cuts as applied to the data.
The simulations describe the data reasonably well.
The data distributions were then
corrected for the effects of selection cuts,
detector acceptance, efficiency, and
resolution, particle decays and interactions
within the detector, and for initial state photon radiation
using bin-by-bin correction factors determined from the Monte Carlo
simulation. The correction factors $C_D$
are shown in Figs. 10(b)--12(b);
the errors comprise the sum in quadrature of the
statistical component from the finite size of the Monte Carlo event
sample, and the systematic uncertainty derived as described in the
previous section.
The hadron level
data are listed in Tables VIII--X, together with statistical
and systematic errors; the central values represent the data corrected
by the central values of the correction factors.
 
The data were further corrected bin-by-bin for the effects of
hadronisation. The hadronisation correction factors are shown in
Figs. 10(c)--12(c); the errors comprise the sum in quadrature of the
statistical component from the finite size of the Monte Carlo event
sample, and the systematic uncertainty.
The fully-corrected data are shown in Figs.~10(a)--12(a); the data points
correspond to the central values of the correction factors,
and the errors shown comprise the
statistical and total systematic components added in quadrature.
Also shown in Figs.~10(a)--12(a) are the parton-level predictions of
the JETSET 7.4
$\Oa$ matrix element, $\Oaa$ matrix element, and parton shower
options, and the HERWIG 5.7 parton shower. All calculations describe
the data well, and higher-order corrections to the $\Oa$ predictions
are seen to be small.
 
The data were divided into four samples according to the thrust
values of the events: (i) $0.70< T < 0.80$, (ii) $0.80< T < 0.85$,
(iii) $0.85< T < 0.90$ and (iv) $0.90< T < 0.95$. The distributions of
cos$\theta$, cos$\theta_N$ and $\chi$ are shown for these four
ranges in Figs.~13, 14 and 15 respectively. Also shown in these figures
are fits of Eqs.~(7), (8) and (9) (Section 2), where the parameters
$\alpha(T)$, $\alpha_N(T)$ and $\beta(T)$ were determined, respectively,
from the fits. The
fitted values of these parameters are listed in Table XI, and are
shown in Fig.~16, where they are compared with the leading-order
QCD predictions and with the
predictions of the scalar and tensor gluon models. Values of $\chi^2$
for these comparisons are given in Table XII.
The data are in agreement with
the QCD predictions, and the scalar and tensor gluon predictions are
disfavoured. It should be noted, however,
that the event plane orientation angle distributions
are less sensitive to the different gluon spin cases
than are the jet energy distributions discussed in the previous section.
 
\section{Conclusions}
 
We have measured distributions of the jet energies, and of the
orientation angles
of the event plane, in \ep \ra \z0 \ra three-jet events recorded in the
SLD experiment at SLAC.
Our measurements of these quantities are consistent with those from
other experiments
\cite{LTHREE,OPAL,DELPHI} at the \z0 resonance.
We have compared our measurements with QCD predictions and with models
of strong interactions incorporating scalar or tensor gluons; this
represents the first comparison with a tensor gluon calculation.
 
The leading-order vector gluon (QCD) calculation describes the basic
shape of the scaled jet energy
distributions, and addition of higher-order perturbative
contributions leads to a reasonable description of the finer details of
these distributions, provided the regions of phase space are avoided
where soft and collinear singularities need to be resummed.
One may speculate that the
addition of as yet uncalculated higher-order QCD
contributions may yield further improvement. The shapes of the
jet energy distributions cannot be described by leading-order
models incorporating either scalar or tensor gluons alone.
However, the {\it ad hoc} addition
of leading-order contributions from scalar and tensor gluons,
each with arbitrary relative weight, to the
QCD predictions can also improve the description of the data;
even for the QCD parton shower calculations slightly better fit
qualities are obtained with such contributions included.
The allowed relative contributions of scalar and tensor gluons depend
upon the order of the vector calculation, as well as the observable;
the smallest allowed
contribution of 0.1\% for both scalar and tensor gluons
is obtained with the HERWIG parton shower fit to the scaled energy of
the most energetic jet.
 
The event plane orientation angles are well described by $\Oa$ QCD
and higher-order corrections are small. These quantities
are less sensitive to the gluon spin than the jet energies, but the
data disfavor the scalar and tensor hypotheses.
 
\section{Acknowledgements}
 
We thank Lance Dixon for contributions to the tensor gluon
model. We thank the personnel of the SLAC accelerator department
and the technical staffs of our collaborating institutions for their
efforts which resulted in the successful operation of the SLC and the
SLD.
 
\section*{Appendix: Tensor Gluon Model}
 
Since the tensor gluon toy model is new, whereas the
vector and scalar cases have been studied in detail in the
literature, we discuss briefly how Eq.~6 was obtained.
 
The only well-known theory involving the exchange of massless, spin-2
gauge fields is the quantized version of General
Relativity, which is both highly non-linear and non-renormalizable.
To obtain a simple parallel model for tensor gluons,
which couple only to color non-singlet sources, we
begin by linearizing the theory of quantum gravity based on General
Relativity by keeping only the
lowest order terms in the coupling and by ignoring the tensor field
self-interactions \cite{cjss}. Although now linear, the theory remains
non-renormalizable, as will be the tensor gluon model,
which should be viewed only as a
toy model against which to test the predictions of QCD.
 
If tensor gluons behaved in the same way as
gravitons one could write down the complete gauge-invariant
amplitude for the tree-level process
\z0 \ra \qqg. The various contributions arise from a set
of four Feynman diagrams: the usual two
which involve gluon bremsstrahlung from the q or $\overline{\rm q}$
in the final
state, the bremsstrahlung of a tensor gluon from the \z0 in the initial
state, producing an off-shell \z0 which
`decays' to q$\overline{\rm q}$,
and finally a new $Z^0$\qqg contact interaction.
We need to remove or modify the $Z^0 Z^0$g piece of the amplitude as
the \z0 is known phenomenologically not to carry a color charge.
 
We consider two possible approaches to this problem. In the first
instance we surrender the possibility of a gauge symmetry for the
tensor gluon theory and omit the diagram involving the $Z^0 Z^0$g
vertex. (We note that the scalar gluon model is also not
a gauge theory.) In this case, using the Feynman gauge for the tensor
gluon, we arrive at the distribution given
in Eq. 6. A second possibility is to mimic the quantum gravity theory as
far as possible and include the $Z^0 Z^0$g diagram
in a modified form. To do this we extend the particle spectrum of
the Standard Model by introducing a color-octet partner to the
$Z^0$, $Z^0_8$, which is degenerate with the $Z^0$ and couples to quarks
in exactly the same way as does the $Z^0$,
except for the presence of color generators. The problematic
$Z^0 Z^0$g vertex is now replaced by the $Z^0 Z^0_8\,$g coupling.
In this case we arrive at a form for the tensor distribution
given by \cite{LD}:
$$
 {1\over\sigma}{d^2\sigma^T\over d x_1d x_2} \quad \propto \quad
{{(x_1+x_2-1)(x_1^2 + x_2^2)}
\over{(2- x_1- x_2)^2}}\quad +
\quad\quad\quad
\quad\quad\quad
\quad\quad\quad
\quad\quad\quad
$$
\begin{eqnarray}
{{(1-x_2)(x_1^2 + (2-x_1-x_2)^2)}
\over{x_2^2}} \quad + \quad
{{(1-x_1)(x_2^2 + (2-x_1-x_2)^2)}
\over{x_1^2}},
\end{eqnarray}
which, apart from the overall normalisation, is the same as that for
graviton radiation in \z0 decays.
Although algebraically different, this form yields
numerically similar results to Eq.~6 (Fig.~17).
 
In the analysis presented in the text the comparison of the tensor
model with the data is based on Eq.~6. It is clear from
Fig.~17, however, that our conclusions would not differ
if Eq.~18 had been chosen instead.
 
 
\vfill
\eject

\vfill\eject
 
 
\begin{table}[t]
 
 
\begin{center}
\renewcommand{\arraystretch}{0.8}
 
\begin{tabular}{|c||c|c|c|}  \hline
$x_1$ & $\frac{1}{\sigma_{3-jet}}\frac{d\sigma}{dx_1}$ & stat. &
exp.~syst. \\ \hline
0.676     & 0.025      &  0.007   & 0.008    \\
0.700     & 0.072      &  0.016   & 0.018    \\
0.724     & 0.133      &  0.018   & 0.022    \\
0.748     & 0.260      &  0.025   & 0.033    \\
0.772     & 0.423      &  0.028   & 0.044    \\
0.796     & 0.530      &  0.032   & 0.044    \\
0.820     & 0.749      &  0.039   & 0.048    \\
0.844     & 1.065      &  0.048   & 0.061    \\
0.868     & 1.603      &  0.056   & 0.071    \\
0.892     & 2.351      &  0.069   & 0.088    \\
0.916     & 3.83       &  0.09    & 0.11     \\
0.940     & 6.74       &  0.11    & 0.14     \\
0.964     & 13.80      &  0.17    & 0.27     \\
0.988     & 9.08       &  0.13    & 0.17     \\  \hline
\end{tabular}
 
\end{center}
\normalsize
 
Table I.
The measured scaled jet energy of the highest-energy jet in 3-jet events.
The data were corrected for detector effects and
for initial state photon radiation.
The first error is statistical, and
the second represents the experimental systematic uncertainty.
 
\end{table}
 
\begin{table}[t]
 
 
\begin{center}
\renewcommand{\arraystretch}{0.8}
 
\begin{tabular}{|c||c|c|c|}  \hline
$x_2$ & $\frac{1}{\sigma_{3-jet}}\frac{d\sigma}{dx_2}$ & stat. &
exp.~syst. \\ \hline
0.5275    & 0.490      &  0.024   & 0.031    \\
0.5625    & 1.031      &  0.039   & 0.050    \\
0.5975    & 1.267      &  0.043   & 0.050    \\
0.6325    & 1.356      &  0.044   & 0.051    \\
0.6675    & 1.546      &  0.048   & 0.058    \\
0.7025    & 1.689      &  0.048   & 0.057    \\
0.7375    & 1.815      &  0.051   & 0.068    \\
0.7725    & 1.938      &  0.053   & 0.061    \\
0.8075    & 2.089      &  0.055   & 0.063    \\
0.8425    & 2.619      &  0.060   & 0.071    \\
0.8775    & 2.966      &  0.063   & 0.074    \\
0.9125    & 3.391      &  0.064   & 0.082    \\
0.9475    & 3.813      &  0.062   & 0.079    \\
0.9825    & 2.205      &  0.056   & 0.075    \\  \hline
\end{tabular}
 
\end{center}
\normalsize
 
Table II.
The measured scaled jet energy of the second
highest-energy jet in 3-jet events.
The data were corrected for detector effects and
for initial state photon radiation.
The first error is statistical, and
the second represents the experimental systematic uncertainty.
 
\end{table}
 
\clearpage
 
\begin{table}[t]
 
\begin{center}
\renewcommand{\arraystretch}{0.8}
 
\begin{tabular}{|c||c|c|c|}  \hline
$x_3$ & $\frac{1}{\sigma_{3-jet}}\frac{d\sigma}{dx_3}$ & stat. &
exp.~syst. \\ \hline
0.0225    & 1.095      &  0.037   & 0.050    \\
0.0675    & 2.622      &  0.044   & 0.059    \\
0.1125    & 2.632      &  0.048   & 0.069    \\
0.1575    & 2.340      &  0.049   & 0.060    \\
0.2025    & 2.228      &  0.049   & 0.060    \\
0.2475    & 1.878      &  0.046   & 0.054    \\
0.2925    & 1.645      &  0.043   & 0.052    \\
0.3375    & 1.502      &  0.040   & 0.051    \\
0.3825    & 1.386      &  0.040   & 0.049    \\
0.4275    & 1.400      &  0.039   & 0.048    \\
0.4725    & 1.356      &  0.038   & 0.045    \\
0.5175    & 1.090      &  0.035   & 0.043    \\
0.5625    & 0.378      &  0.022   & 0.028    \\
0.6075    & 0.188      &  0.016   & 0.022    \\
0.6525    & 0.037      &  0.008   & 0.009    \\  \hline
\end{tabular}
 
\end{center}
\normalsize
 
Table III.
The measured scaled jet energy of the lowest-energy jet in 3-jet events.
The data were corrected for detector effects and
for initial state photon radiation.
The first error is statistical, and
the second represents the experimental systematic uncertainty.
 
\end{table}
 
\begin{table}[t]
 
\begin{center}
\renewcommand{\arraystretch}{0.8}
 
\begin{tabular}{|c||c|c|c|}  \hline
cos$\theta_{EK}$ &
$\frac{1}{\sigma_{3-jet}}\frac{d\sigma}{dcos\theta_{EK}}$
& stat. & exp.~syst. \\ \hline
0.025     & 0.689      &  0.028   & 0.032    \\
0.075     & 0.692      &  0.028   & 0.032    \\
0.125     & 0.678      &  0.027   & 0.035    \\
0.175     & 0.669      &  0.027   & 0.032    \\
0.225     & 0.671      &  0.026   & 0.030    \\
0.275     & 0.716      &  0.027   & 0.031    \\
0.325     & 0.718      &  0.026   & 0.034    \\
0.375     & 0.733      &  0.028   & 0.043    \\
0.425     & 0.819      &  0.028   & 0.034    \\
0.475     & 0.803      &  0.029   & 0.037    \\
0.525     & 0.835      &  0.029   & 0.035    \\
0.575     & 0.906      &  0.030   & 0.036    \\
0.625     & 1.055      &  0.032   & 0.038    \\
0.675     & 1.207      &  0.034   & 0.047    \\
0.725     & 1.290      &  0.034   & 0.041    \\
0.775     & 1.420      &  0.035   & 0.047    \\
0.825     & 1.507      &  0.035   & 0.056    \\
0.875     & 1.700      &  0.035   & 0.043    \\
0.925     & 1.696      &  0.032   & 0.043    \\
0.975     & 0.776      &  0.029   & 0.039    \\  \hline
\end{tabular}
 
\end{center}
\normalsize
 
Table IV.
The measured Ellis-Karliner angle distribution in 3-jet events.
The data were corrected for detector effects and
for initial state photon radiation.
The first error is statistical, and
the second represents the experimental systematic uncertainty.
 
\end{table}
 
\clearpage
 
\begin{table}[t]
 
\begin{center}
\renewcommand{\arraystretch}{0.8}
 
\begin{tabular}{|l|c|c|c|}  \hline
Parameter       & Variable Name  & Default  & Optimised  \\ \hline
$\Lambda_{QCD}$ & PARJ(81) & 0.29 GeV & 0.26 GeV         \\
$\sigma_q$      & PARJ(21) & 0.36 GeV/$c$ & 0.39 GeV/$c$ \\
$a$             & PARJ(41) & 0.3      & 0.18             \\
$b$             & PARJ(42) & 0.58 GeV$^{-2}$ & 0.34 GeV$^{-2}$ \\
$\epsilon_c$    & PARJ(54) & $-0.05$  & $-0.06$          \\
$\epsilon_b$    & PARJ(55) & $-0.005$ &$-0.006$          \\
diquark prob.\hfill & PARJ(1) &  0.10     & 0.08         \\
s quark prob.\hfill & PARJ(2) &  0.30     & 0.28         \\
s diquark prob.\hfill & PARJ(3)  & 0.40   & 0.60         \\
V meson prob. (u,d) \hfill & PARJ(11)  & 0.50    & 0.50  \\
V meson prob. (s)\hfill &PARJ(12) & 0.60     & 0.45      \\
V meson prob. (c,b) \hfill & PARJ(13)  & 0.75    &  0.53 \\
$\eta$' prob.\hfill & PARJ(26)  & 0.40     & 0.20        \\ \hline
\end{tabular}
 
\end{center}
\normalsize
 
Table V.
Parameters in JETSET 7.4 that were
changed from default values (see text).
 
\end{table}
 
\begin{table}[t]
 
\begin{center}
\renewcommand{\arraystretch}{0.8}
 
\begin{tabular}{|l|c|c|c|c|c|}  \hline
Distribution & \# bins &
JETSET $\Oa$ & JETSET $\Oaa$ & JETSET PS & HERWIG PS      \\ \hline
$x_1$             & 14 (12) & 88.2 (72.6) & 38.5 (26.3) & 13.5 (6.3)
& 11.2 (10.6) \\
$x_2$             & 14 (12) & 37.8 (20.0) & 36.8 (12.2) & 34.9 (21.0)
& 15.2 (6.5) \\
$x_3$             & 15 (13) & 92.9 (49.8) & 86.5 (29.6) & 22.3 (17.5)
& 25.7 (11.8) \\
cos$\theta_{EK}$  & 20 (18) & 60.6 (26.3) & 86.2 (44.6) & 15.8 (9.0)
& 48.2 (30.2) \\ \hline
\end{tabular}
 
\end{center}
\normalsize
 
Table VI. Numbers of bins and
$\chi^2$ values for comparison between fully corrected data and
parton-level QCD Monte Carlo calculations. Values in parentheses
are for the restricted ranges which exclude the regions where soft and
collinear contributions are expected to be large.
 
\end{table}
 
\begin{table}[t]
 
\begin{center}
\renewcommand{\arraystretch}{0.8}
 
\begin{tabular}{|l|c|c|c|c|}  \hline
Distribution & \# bins & Vector & Scalar & Tensor \\ \hline
$x_1$        & 12      & 45.2   & 1116.4 & 141.9  \\
$x_2$        & 12      & 33.5   & 1321.7 & 490.6  \\
$x_3$        & 13      & 39.9   & 2011.4 & 546.9  \\
cos$\theta_{EK}$ & 18  & 19.5   & 1684.0 & 772.1  \\ \hline
\end{tabular}
 
\end{center}
\normalsize
 
Table VII. Numbers of bins and
$\chi^2$ values for comparison between fully corrected data and
leading-order vector (QCD), scalar, and tensor gluon calculations.
 
\end{table}
 
\clearpage

\begin{table}[t]
 
\begin{center}
\renewcommand{\arraystretch}{0.8}

\begin{tabular}{|c||c|c|c|}  \hline
cos$\theta$ & $\frac{1}{\sigma_{3-jet}}\frac{d\sigma}{dcos\theta}$
& stat. & exp.~syst. \\ \hline
0.071     & 0.792      &  0.021   & 0.031    \\
0.214     & 0.822      &  0.023   & 0.031    \\
0.357     & 0.853      &  0.023   & 0.030    \\
0.500     & 0.982      &  0.024   & 0.033    \\
0.643     & 1.088      &  0.026   & 0.031    \\
0.786     & 1.135      &  0.028   & 0.035    \\
0.929     & 1.306      &  0.035   & 0.090    \\  \hline
\end{tabular}
 
\end{center}
\normalsize

Table VIII.
The measured polar angle w.r.t. the electron beam
of the highest-energy jet in 3-jet events.
The data were corrected for detector effects and
for initial state photon radiation. The first error is statistical, and
the second represents the experimental systematic uncertainty.
 
\end{table}
 
\begin{table}[t]
 
\begin{center}
\renewcommand{\arraystretch}{0.8}

\begin{tabular}{|c||c|c|c|}  \hline
cos$\theta_N$ & $\frac{1}{\sigma_{3-jet}}\frac{d\sigma}{dcos\theta_N}$
& stat. & exp.~syst. \\ \hline
0.071     & 1.159      &  0.034   & 0.076    \\
0.214     & 1.079      &  0.029   & 0.046    \\
0.357     & 1.110      &  0.026   & 0.029    \\
0.500     & 0.969      &  0.025   & 0.028    \\
0.643     & 0.967      &  0.025   & 0.035    \\
0.786     & 0.917      &  0.023   & 0.036    \\
0.929     & 0.804      &  0.020   & 0.030    \\  \hline
\end{tabular}
 
\end{center}
\normalsize
 
Table IX.
The measured polar angle w.r.t. the electron beam
of the normal to the three-jet plane.
The data were corrected for detector effects and
for initial state photon radiation. The first error is statistical, and
the second represents the experimental systematic uncertainty.
 
\end{table}
 
\begin{table}[t]
 
\begin{center}
\renewcommand{\arraystretch}{0.8}

\begin{tabular}{|c||c|c|c|}  \hline
$\chi$ (rad.) & $\frac{1}{\sigma_{3-jet}}\frac{d\sigma}{d\chi}$
& stat. & exp.~syst. \\ \hline
0.112     & 0.671      &  0.025   & 0.034    \\
0.336     & 0.644      &  0.025   & 0.027    \\
0.561     & 0.633      &  0.025   & 0.026    \\
0.785     & 0.642      &  0.024   & 0.025    \\
1.009     & 0.635      &  0.023   & 0.025    \\
1.234     & 0.592      &  0.021   & 0.023    \\
1.458     & 0.645      &  0.021   & 0.023    \\ \hline
\end{tabular}
 
\end{center}
\normalsize
 
Table X.
The measured angle between the event plane and the plane
containing the highest-energy jet and the electron beam.
The data were corrected for detector effects and
for initial state photon radiation. The first error is statistical, and
the second represents the experimental systematic uncertainty.
 
\end{table}
 
\clearpage
 

\begin{table}[t]
 
\begin{center}
\renewcommand{\arraystretch}{0.8}

\begin{tabular}{|c||c|c||c|c||c|c|}  \hline
Thrust range & $\alpha(T)$ & $\chi^2$ & $\alpha_N(T)$ & $\chi^2$ &
$\beta(T)$ & $\chi^2$ \\ \hline
$0.7<T<0.8$  & $0.61\pm0.18$ & 6.1 & $-0.42\pm0.10$ & 1.9
             & $0.090\pm0.069$ & 5.4 \\
$0.8<T<0.85$ & $0.83\pm0.19$ & 3.6 & $-0.31\pm0.11$ & 0.6
             & $0.034\pm0.071$ & 3.3 \\
$0.85<T<0.9$ & $0.82\pm0.12$ & 8.3 & $-0.33\pm0.07$ & 7.8
             & $0.004\pm0.041$ & 4.4 \\
$0.9<T<0.95$ & $0.81\pm0.09$ & 2.6 & $-0.26\pm0.06$ & 6.8
             & $-0.033\pm0.030$ & 0.5 \\ \hline
\end{tabular}
 
\end{center}
\normalsize
 
Table XI.
Thrust ranges, values and errors of the fit parameters $\alpha$,
$\alpha_N$ and $\beta$, and $\chi^2$ values for the fits. For each
fitted observable there are 7 bins.
 
\end{table}
 
\begin{table}[t]
 
\begin{center}
\renewcommand{\arraystretch}{0.8}

\begin{tabular}{|c||c|c|c|}  \hline
Gluon spin   & $\alpha(T)$ & $\alpha_N(T)$ & $\beta(T)$  \\ \hline
Vector       & 3.0         & 2.8           & 2.4 \\
Scalar       & 17.4        & 38.0          & 8.8 \\
Tensor       & 7.3         & 5.7           & 4.4 \\ \hline
\end{tabular}
 
\end{center}
\normalsize
 
Table XII.
Values of $\chi^2$ for comparisons between the predictions including
vector, scalar or tensor gluons for the coefficients
$\alpha(T)$, $\alpha_N(T)$ and $\beta(T)$ and the measured values
(Fig.~16).
 
\end{table}
 
\clearpage
 
\clearpage
\section*{Figure captions }
 
\noindent
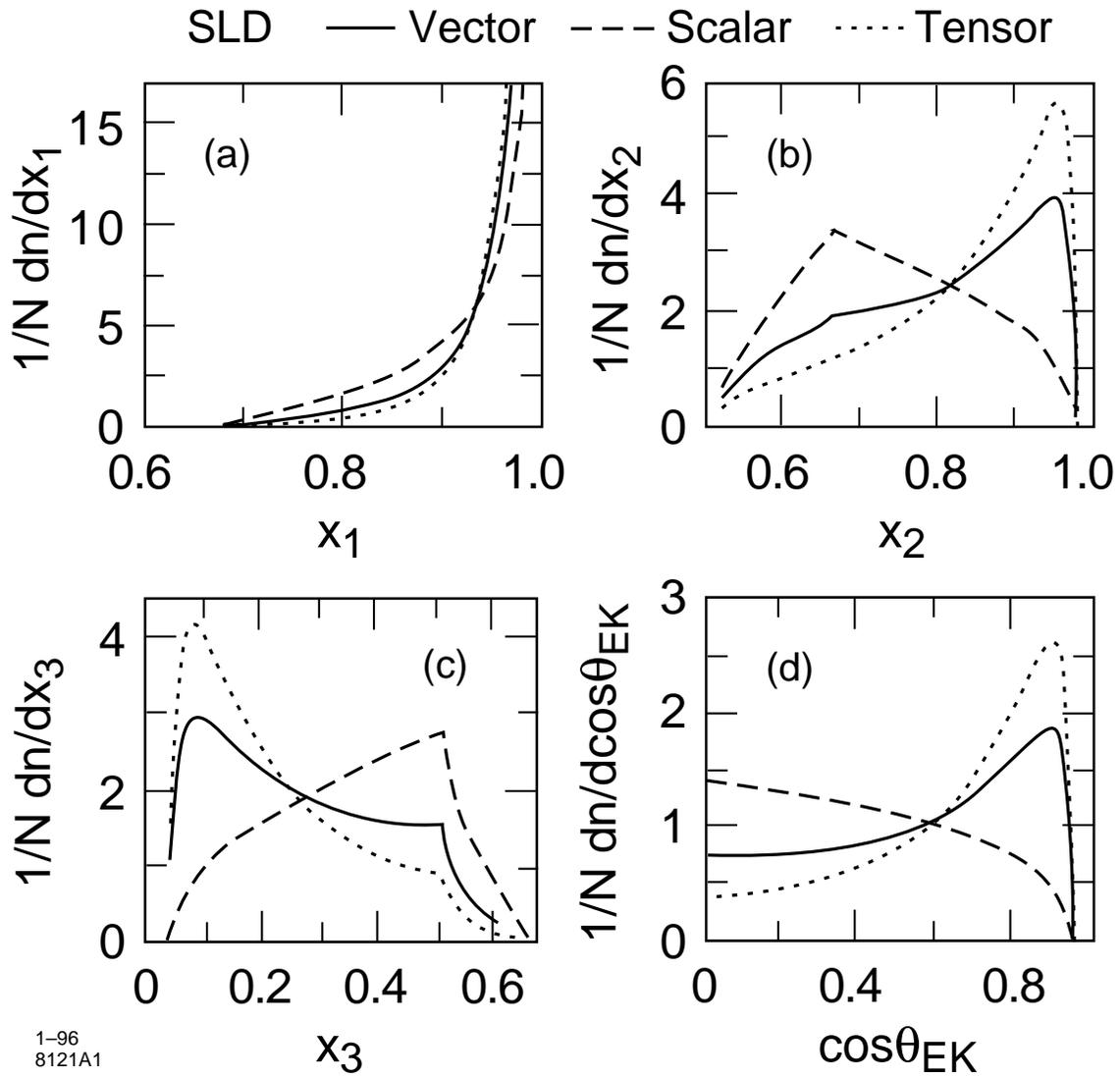
{\bf Figure 1}.
Leading-order calculations, incorporating vector (solid), scalar
(long dashed), and tensor (short dashed) gluons, of distributions of:
(a) scaled energy of the highest-energy jet; (b) scaled energy of the
second highest-energy jet; (c) scaled energy of the lowest-energy
jet; (d) the Ellis-Karliner angle.
 
\noindent
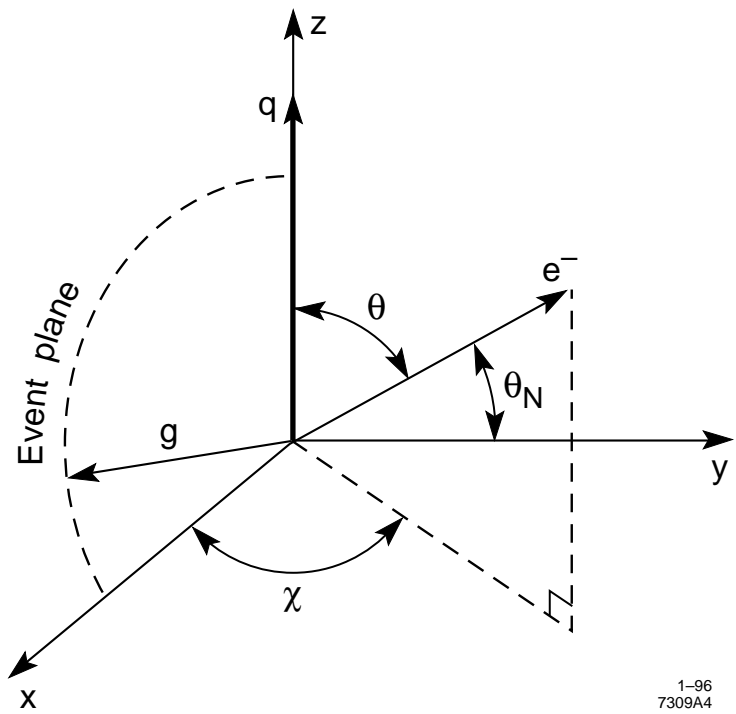
{\bf Figure 2}.
Definition of the Euler angles $\theta$, $\theta_N$ and $\chi$ that
decribe the orientation of the event plane.
 
\noindent
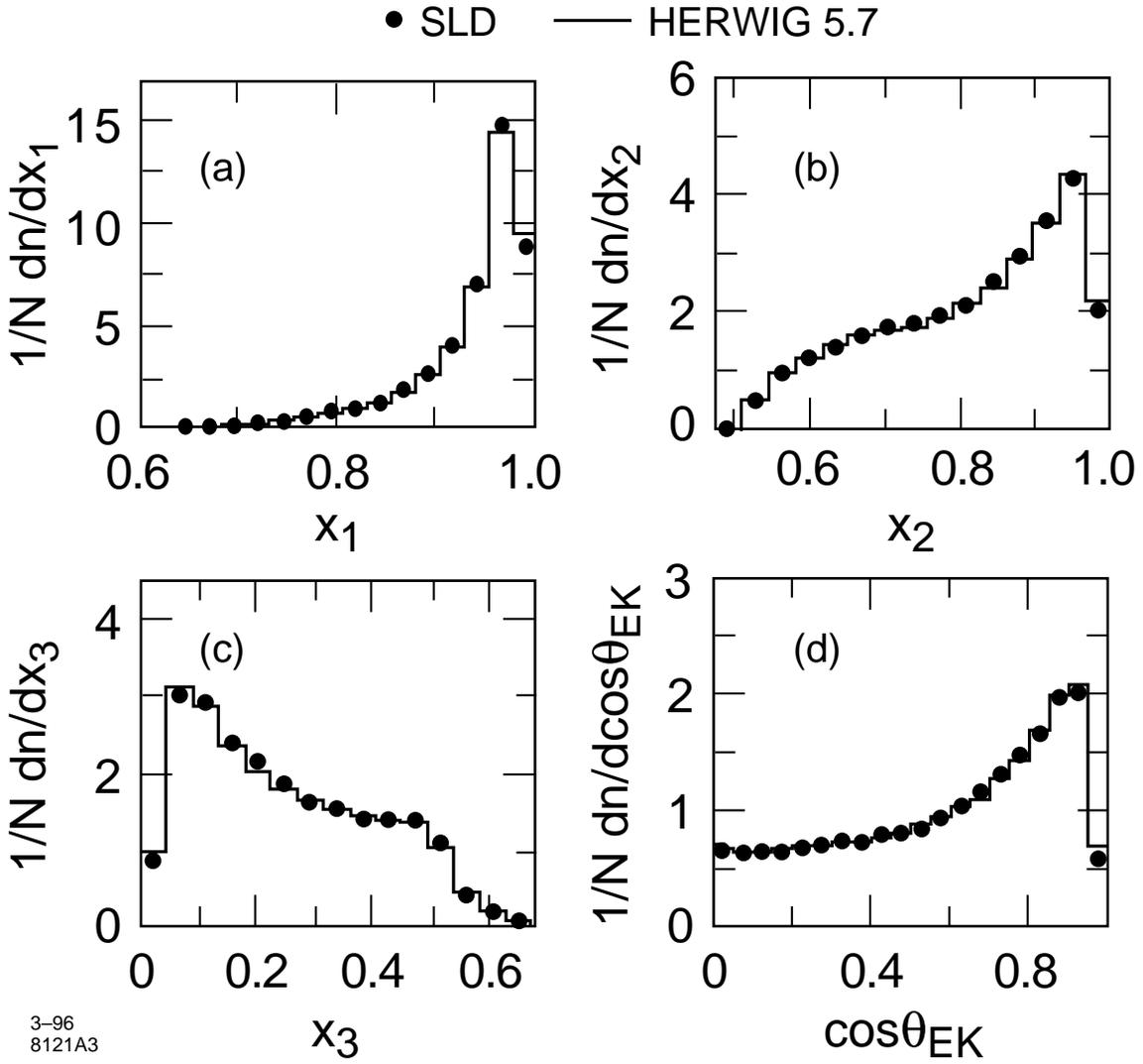
{\bf Figure 3}.
Measured distributions (dots) of:
(a) scaled energy of the highest-energy jet; (b) scaled energy of the
second highest-energy jet; (c) scaled energy of the lowest-energy
jet; (d) the Ellis-Karliner angle. The errors are statistical only.
The predictions of a Monte Carlo simulation are
shown as solid histograms.
 
\noindent
{\bf Figure 4}.
(a) The measured distribution (dots) of
the scaled energy of the highest-energy
jet, fully-corrected to the parton
level, compared with QCD Monte Carlo calculations.
The errors comprise the total statistical and systematic components
added in quadrature.
The correction factors for detector effects and initial-state
radiation (b) and for hadronisation effects (c);
the inner error bars show the statistical
component and the outer error bars the total uncertainty.
 
\noindent
{\bf Figure 5}.
(a) The measured distribution (dots) of
the scaled energy of the second highest-energy jet,
fully-corrected to the parton
level, compared with QCD Monte Carlo calculations.
The errors comprise the total statistical and systematic components
added in quadrature.
The correction factors for detector effects and initial-state
radiation (b) and for hadronisation effects (c);
the inner error bars show the statistical
component and the outer error bars the total uncertainty.
 
\noindent
{\bf Figure 6}.
(a) The measured distribution (dots) of
the scaled energy of the lowest-energy jet, fully-corrected to the parton
level, compared with QCD Monte Carlo calculations.
The errors comprise the total statistical and systematic components
added in quadrature.
The correction factors for detector effects and initial-state
radiation (b) and for hadronisation effects (c);
the inner error bars show the statistical
component and the outer error bars the total uncertainty.
 
\noindent
{\bf Figure 7}.
(a) The measured distribution (dots) of
the Ellis-Karliner angle, fully-corrected to the parton
level, compared with QCD Monte Carlo calculations.
The errors comprise the total statistical and systematic components
added in quadrature.
The correction factors for detector effects and initial-state
radiation (b) and for hadronisation effects (c);
the inner error bars show the statistical
component and the outer error bars the total uncertainty.
 
\noindent
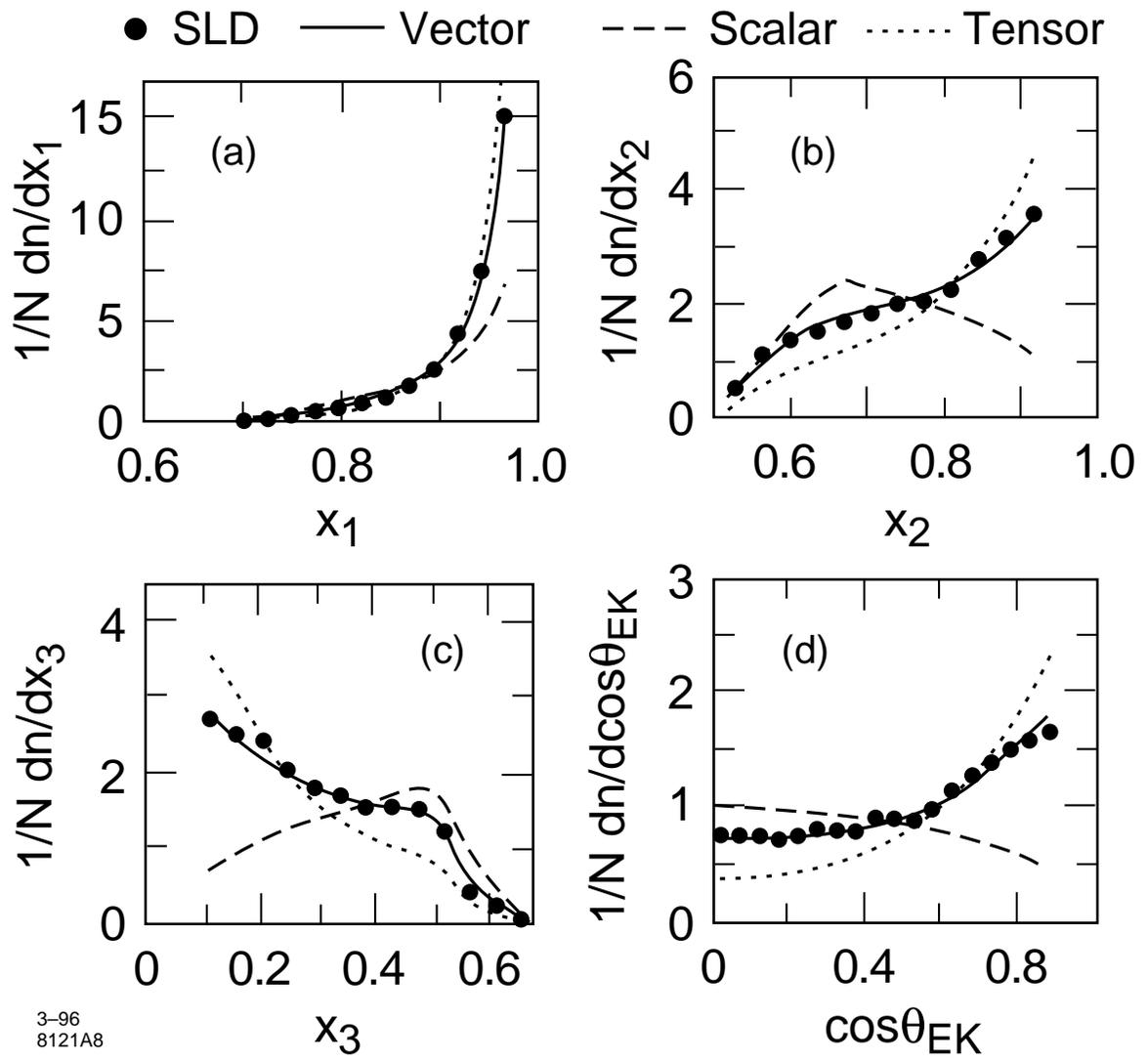
{\bf Figure 8}.
Measured distributions, fully corrected to the parton level (dots), of:
(a) scaled energy of the highest-energy jet;
(b) scaled energy of the second highest-energy jet;
(c) scaled energy of the lowest-energy jet; (d) the
Ellis-Karliner angle.
The errors comprise the total statistical and systematic components
added in quadrature.
The leading-order predictions described in
Section 2 are shown as lines: vector (solid), scalar (long dashed),
and tensor (short dashed).
 
\noindent
{\bf Figure 9}.
Measured distributions (dots) of the event plane orientation angles:
(a) cos$\theta$, (b) cos$\theta_N$, (c) $\chi$.
The errors are statistical only.
The predictions of a Monte Carlo simulation are
shown as solid histograms.
 
\noindent
{\bf Figure 10}.
(a) The measured distribution (dots) of
cos$\theta$, fully-corrected to the parton
level, compared with QCD Monte Carlo calculations.
The errors comprise the total statistical and systematic components
added in quadrature.
The correction factors for detector effects and initial-state
radiation (b) and for hadronisation effects (c);
the inner error bars show the statistical
component and the outer error bars the total uncertainty.
 
\noindent
{\bf Figure 11}.
(a) The measured distribution (dots) of
cos$\theta_N$, fully-corrected to the parton
level, compared with QCD Monte Carlo calculations.
The errors comprise the total statistical and systematic components
added in quadrature.
The correction factors for detector effects and initial-state
radiation (b) and for hadronisation effects (c);
the inner error bars show the statistical
component and the outer error bars the total uncertainty.
 
\noindent
{\bf Figure 12}.
(a) The measured distribution (dots) of
$\chi$, fully-corrected to the parton
level, compared with QCD Monte Carlo calculations.
The errors comprise the total statistical and systematic components
added in quadrature.
The correction factors for detector effects and initial-state
radiation (b) and for hadronisation effects (c);
the inner error bars show the statistical
component and the outer error bars the total uncertainty.
 
\noindent
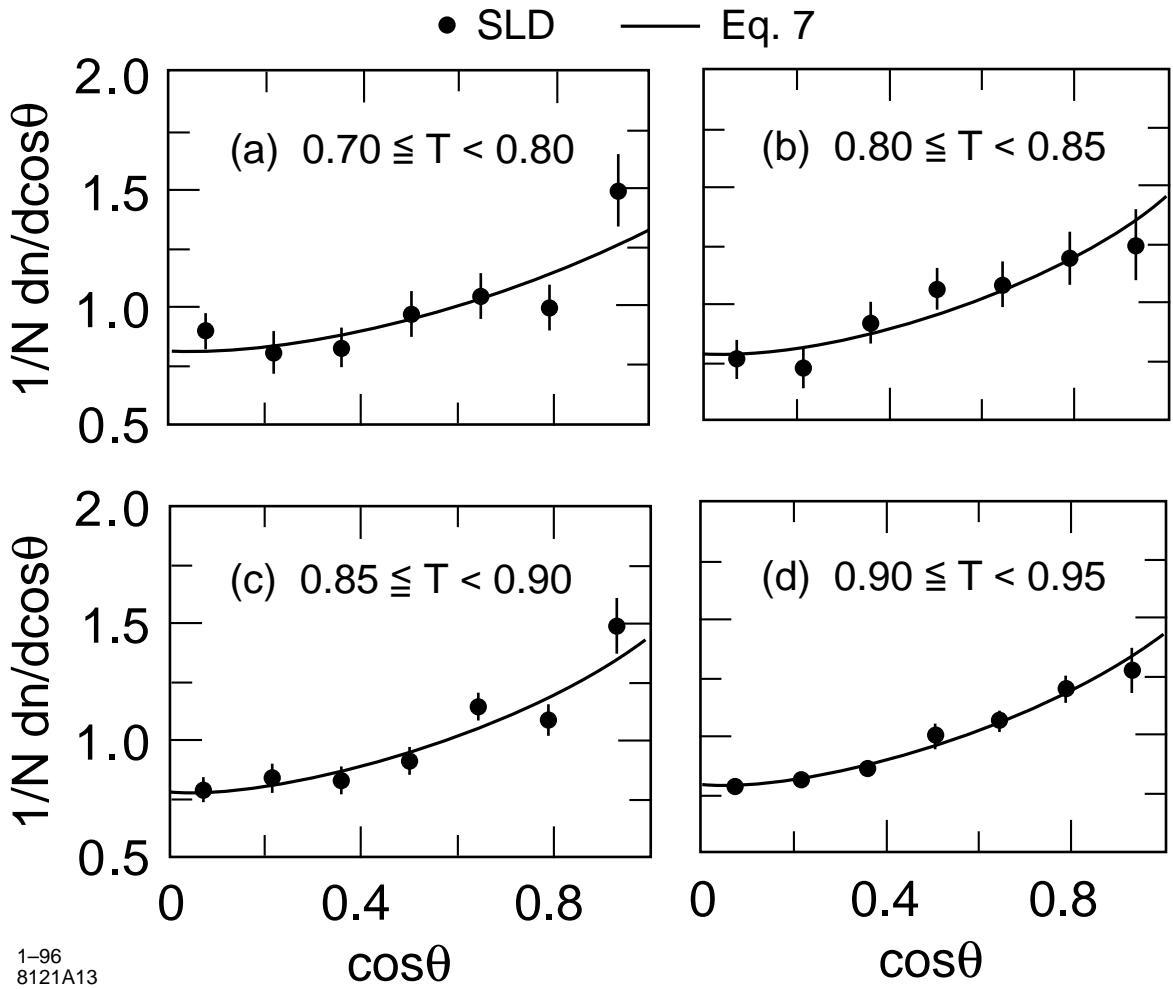
{\bf Figure 13}.
The measured distributions (dots) of
cos$\theta$, fully-corrected to the parton
level, in the event thrust ranges: (a) $0.70< T < 0.80$,
(b) $0.80< T < 0.85$, (c) $0.85< T < 0.90$, (d) $0.90< T < 0.95$.
The errors comprise the total statistical and systematic components
added in quadrature.
Fits of Eq.~7 are shown as solid lines.
 
\noindent
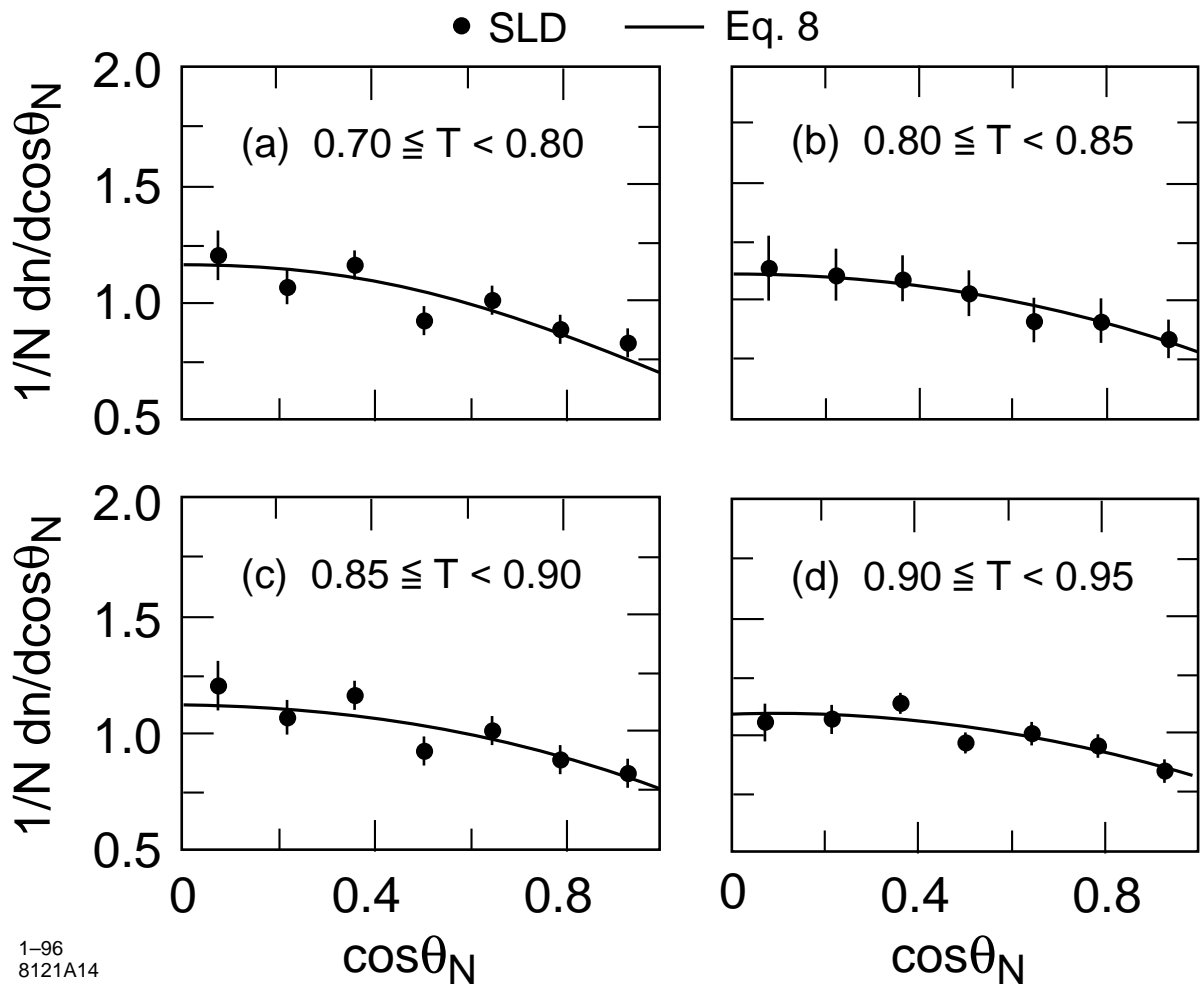
{\bf Figure 14}.
The measured distributions (dots) of
cos$\theta_N$, fully-corrected to the parton
level, in the event thrust ranges: (a) $0.70< T < 0.80$,
(b) $0.80< T < 0.85$, (c) $0.85< T < 0.90$, (d) $0.90< T < 0.95$.
The errors comprise the total statistical and systematic components
added in quadrature.
Fits of Eq.~8 are shown as solid lines.
 
\noindent
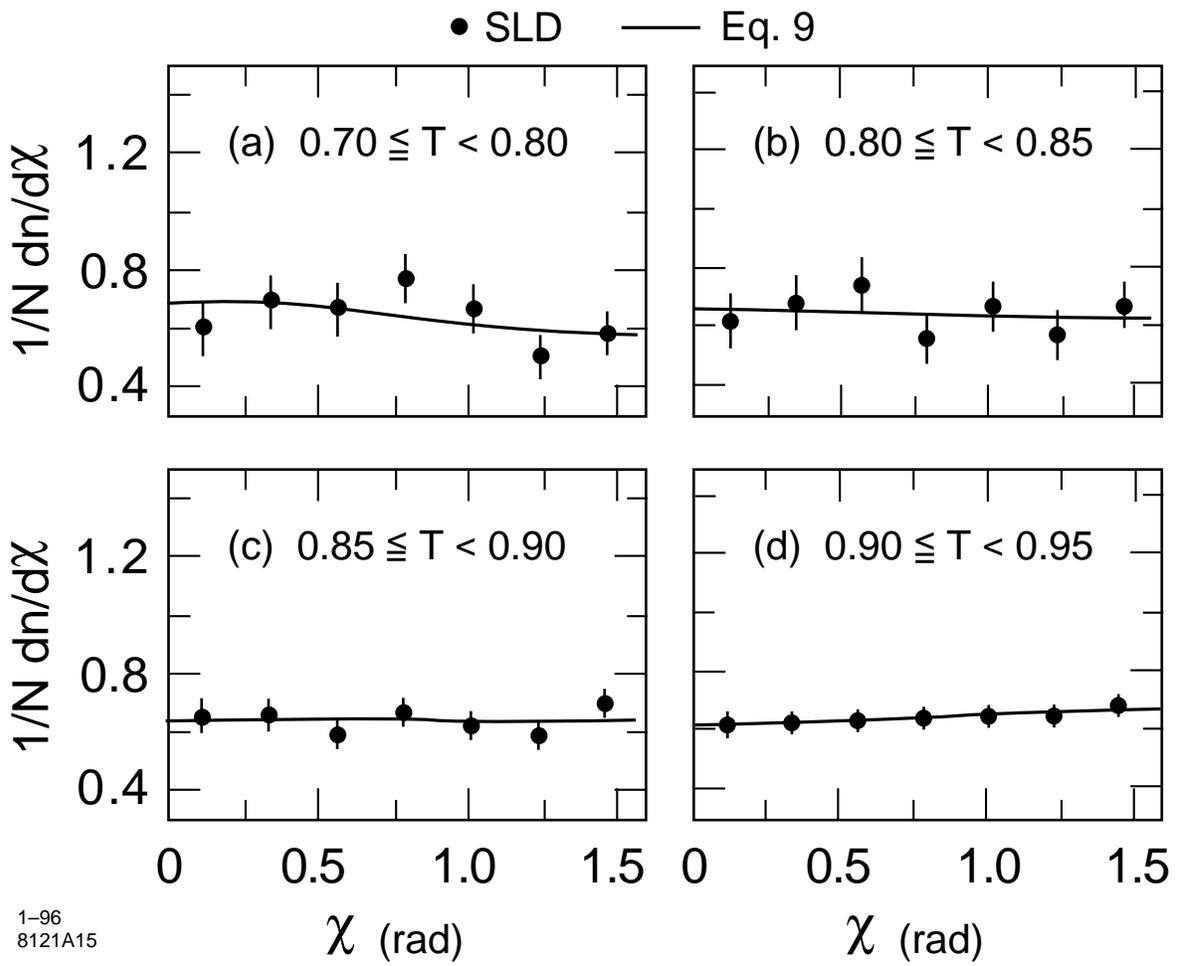
{\bf Figure 15}.
The measured distributions (dots) of
$\chi$, fully-corrected to the parton
level, in the event thrust ranges: (a) $0.70< T < 0.80$,
(b) $0.80< T < 0.85$, (c) $0.85< T < 0.90$, (d) $0.90< T < 0.95$.
The errors comprise the total statistical and systematic components
added in quadrature.
Fits of Eq.~9 are shown as solid lines.
 
\noindent
{\bf Figure 16}.
Coefficients (a) $\alpha(T)$, (b) $\alpha_N(T)$,
(c) $\beta(T)$ from the fits
shown in Figs.~13, 14, 15 respectively.
Also shown are the leading-order vector (solid), scalar
(long dashed) and tensor (short dashed) gluon predictions.
 
\noindent
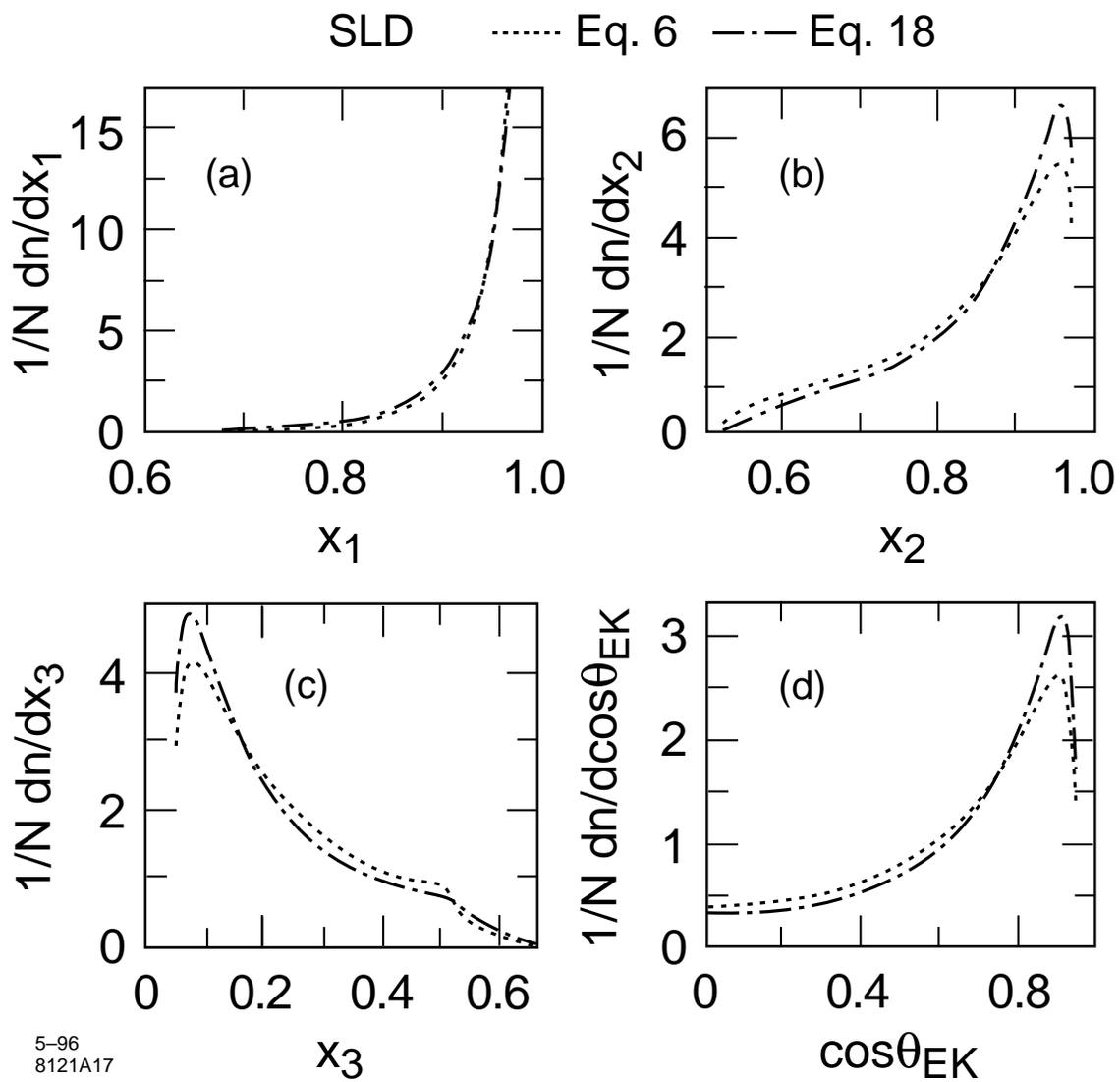
{\bf Figure 17}.
Leading-order tensor gluon model calculations, based on Eq.~6
(short dashed)
and Eq.~18 (dash-dotted), of distributions of:
(a) scaled energy of the highest-energy jet; (b) scaled energy of the
second highest-energy jet; (c) scaled energy of the lowest-energy
jet; (d) the Ellis-Karliner angle.

 
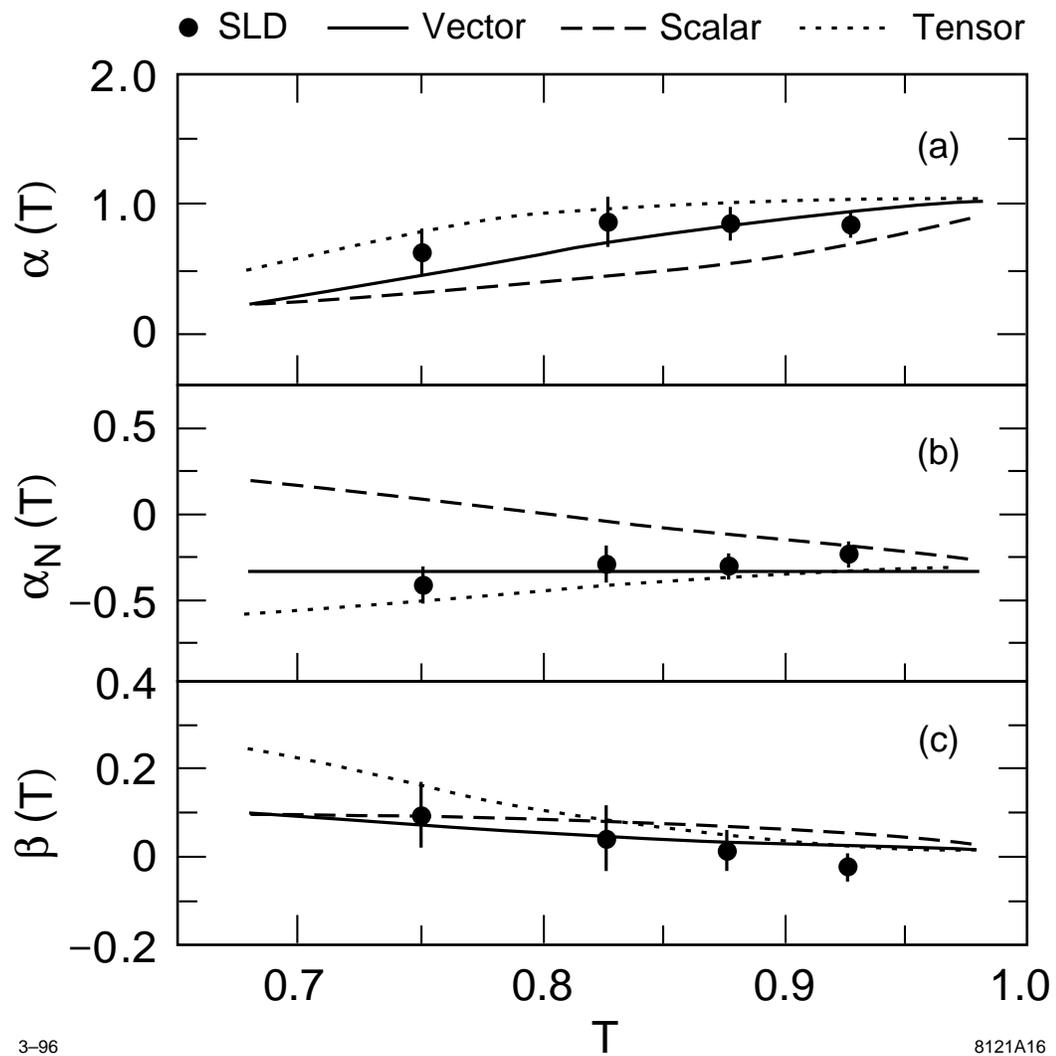
\end{document}

%% file: mydefs.tex
%
%
%
%
 
\catcode`@=11
\def\chkspace{%
  \relax   
  \begingroup\ifhmode\aftergroup\dochksp@ce\fi\endgroup}
\def\dochksp@ce{%
  \unskip              
  \futurelet\chkspct@k\d@chkspc  
}
\def\d@chkspc{%
  \let\nxtsp@ce=\relax
  \ifx\chkspct@k.\else     
    \ifx\chkspct@k,\else
      \ifx\chkspct@k;\else
        \ifx\chkspct@k!\else
          \ifx\chkspct@k?\else
            \ifx\chkspct@k:\else
              \ifx\chkspct@k)\else
              \ifx\chkspct@k(\else
                \ifx\chkspct@k]\else
                  \ifx\chkspct@k-\else
                    \ifx\chkspct@k\egroup\else  
                      \let\nxtsp@ce=\put@space  
                    \fi
                  \fi
                \fi
              \fi
              \fi
            \fi
          \fi
        \fi
      \fi
    \fi
  \fi
  \nxtsp@ce
}
\def\put@space{$\;$}
\catcode`@=12
 
\def\ra{{$\rightarrow$}\chkspace}
\def\etal{{\it et al.}\chkspace}
\def\viz{{\it viz}\chkspace}
\def\adhoc{{\it ad hoc}\chkspace}
\def\ie{{\it i.e.}\chkspace}
\def\ap{{\it a priori}\chkspace}
\def\eg{{\it eg.}\chkspace}
\def\etc{{\it etc.}\chkspace}
\def\ala{{$\grave{a}\; la$}\chkspace}
\def\ibid{{\it ibid}\chkspace}
\def\defacto{{\it de facto}\chkspace}
\def\perse{{\it per se}\chkspace}
\def\apriori{{\it a priori}\chkspace}
 
\def\ep{{e$^+$e$^-$}\chkspace}
\def\epa{{e$^+$e$^-$ annihilation}\chkspace}
\def\mup{{$\mu^+\mu^-$}\chkspace}
\def\taup{{$\tau^+\tau^-$}\chkspace}
\def\qu{\quad}
\def\quu{\quad\quad}
\def\quuuu{\quad\quad\quad\quad}
 
\def\gluino{\relax\ifmmode \tilde{g} \else $\tilde{g}$ \fi\chkspace}
 
\def\qq{q$\overline{\rm q}$\chkspace}
\def\qbar{$\overline{\rm q}$\chkspace}
\def\QQ{Q$\overline{\rm Q}$\chkspace}
\def\pp{p$\overline{\rm p}$\chkspace}
\def\pbar{$\overline{\rm p}$\chkspace}
 
\def\bb{\relax\ifmmode {\rm b}\bar{\rm b}
       \else ${\rm b}\bar{\rm b}$ \fi\chkspace}
\def\cc{\relax\ifmmode {\rm c}\bar{\rm c}
       \else ${\rm c}\bar{\rm c}$ \fi\chkspace}
\def\tt{\relax\ifmmode {\rm t}\bar{\rm t}
       \else ${\rm t}\bar{\rm t}$ \fi\chkspace}
\def\dd{{$d\bar{d}$}\chkspace}
\def\ss{{$s\bar{s}$}\chkspace}
\def\sbar{{$\bar{s}$}\chkspace}
\def\uds{{$u\bar{u},\;d\bar{d},\;s\bar{s}$}\chkspace}
\def\udsc{{$u\bar{u},\;d\bar{d},\;s\bar{s},\;c\bar{c}$}\chkspace}
\def\qqg{\relax\ifmmode {\rm q}\overline{\rm q}{\rm g}
\else q$\overline{\rm q}$g \fi\chkspace}
\def\bbg{{b$\overline{\rm b}$g}\chkspace}
\def\ttg{{t$\overline{\rm t}$g}\chkspace}
\def\QQg{{Q$\overline{\rm Q}$g}\chkspace}
\def\qqgg{{q$\overline{\rm q}$gg}\chkspace}
\def\qqqq{{q$\overline{\rm q}${q$\overline{\rm q}$}}\chkspace}
\def\QQQQ{{Q$\overline{\rm Q}${Q$\overline{\rm Q}$}}\chkspace}
\def\bbbb{{b$\overline{\rm b}${b$\overline{\rm b}$}}\chkspace}
\def\cccc{{c$\overline{\rm c}${c$\overline{\rm c}$}}\chkspace}
 
\def\afb{\relax\ifmmode A_{FB} \else
{{$A_{FB}$}}\fi\chkspace}
\def\afbb{\relax\ifmmode A_{FB}^b \else
{{$A_{FB}^b$}}\fi\chkspace}
\def\pafb{\relax\ifmmode \tilde{A}_{FB} \else
{{$\tilde{A}_{FB}$}}\fi\chkspace}
\def\pafbb{\relax\ifmmode \tilde{A}_{FB}^b \else
{{$\tilde{A}_{FB}^b$}}\fi\chkspace}
 
\def\pafbzo{\relax\ifmmode \tilde{A}_{FB}|_{O(0)} \else
{{$\tilde{A}_{FB}|_{O(0)}$}}\fi\chkspace}
\def\pafbfo{\relax\ifmmode \tilde{A}_{FB}|_{\oalp} \else
{{$\tilde{A}_{FB}|_{\oalp}$}}\fi\chkspace}
\def\pafbso{\relax\ifmmode \tilde{A}_{FB}|_{\oalpsq} \else
{{$\tilde{A}_{FB}|_{\oalpsq}$}}\fi\chkspace}
\def\pafbto{\relax\ifmmode \tilde{A}_{FB}|_{\oalpc} \else
{{$\tilde{A}_{FB}|_{\oalpc}$}}\fi\chkspace}
 
\def\pafbbzo{\relax\ifmmode \tilde{A}_{FB}^b|_{O(0)} \else
{{$\tilde{A}_{FB}^b|_{O(0)}$}}\fi\chkspace}
\def\pafbbfo{\relax\ifmmode \tilde{A}_{FB}^b|_{\oalp} \else
{{$\tilde{A}_{FB}^b|_{\oalp}$}}\fi\chkspace}
\def\pafbbso{\relax\ifmmode \tilde{A}_{FB}^b|_{\oalpsq} \else
{{$\tilde{A}_{FB}^b|_{\oalpsq}$}}\fi\chkspace}
\def\pafbbto{\relax\ifmmode \tilde{A}_{FB}^b|_{\oalpc} \else
{{$\tilde{A}_{FB}^b|_{\oalpc}$}}\fi\chkspace}
 
\def\afbo0{\tilde{A}_{FB}|_{O(0)}}
\def\afbo1{\tilde{A}_{FB}|_{\oalp}}
\def\afbo2{\tilde{A}_{FB}|_{\oalpsq}}
\def\afbo3{\tilde{A}_{FB}|_{\oalpc}}
 
\def\lam{\relax\ifmmode \Lambda_{\overline{MS}}
       \else {{$\Lambda_{\overline{MS}}$}}\fi\chkspace}
\def\lamuds{\relax\ifmmode \Lambda^{(3)}_{\overline{MS}}
       \else {{$\Lambda^{(3)}_{\overline{MS}}$}}\fi\chkspace}
\def\lamudsc{\relax\ifmmode \Lambda^{(4)}_{\overline{MS}}
       \else $\Lambda^{(4)}_{\overline{MS}}$\fi\chkspace}
\def\lamudscb{\relax\ifmmode \Lambda^{(5)}_{\overline{MS}}
       \else $\Lambda^{(5)}_{\overline{MS}}$\fi\chkspace}
\def\alpb{$\alpha_s(b)$\chkspace}
\def\alpc{$\alpha_s(c)$\chkspace}
\def\alpbc{$\alpha_s(bc)$\chkspace}
\def\alpuds{$\alpha_s(uds)$\chkspace}
\def\alpudsc{$\alpha_s(udsc)$\chkspace}
\def\alp{\relax\ifmmode \alpha_s\else $\alpha_s$\fi\chkspace}
\def\alpbar{\relax\ifmmode \overline{\alpha_s}
       \else $\overline{\alpha_s}$\fi\chkspace}
\def\alpmz{\relax\ifmmode \alpha_s(M_Z)\else $\alpha_s(M_Z)$\fi\chkspace}
\def\alpmzsq{\relax\ifmmode \alpha_s(M_Z^2)
       \else $\alpha_s(M_Z^2)$\fi\chkspace}
 
\def\oalp{\relax\ifmmode O(\alpha_s)\else{{O($\alpha_s$)}}\fi\chkspace}
\def\oalpsq{\relax\ifmmode O(\alpha_s^2)
           \else{{O($\alpha_s^2$)}}\fi\chkspace}
\def\oalpc{\relax\ifmmode O(\alpha_s^3)
           \else{{O($\alpha_s^3$)}}\fi\chkspace}
\def\oalpf{\relax\ifmmode O(\alpha_s^4)
           \else{{O($\alpha_s^4$)}}\fi\chkspace}

\def\plb{Phys. Lett.\chkspace}
\def\npb{Nucl. Phys.\chkspace}
\def\rmp{Rev. Mod. Phys.\chkspace}
\def\prl{Phys. Rev. Lett.\chkspace}
\def\prd{Phys. Rev.\chkspace}
\def\zpc{Z. Phys.\chkspace}

\def\z0{{$Z^0$}\chkspace}
\def\Dst{\relax\ifmmode {\rm D}^* \else {D$^*$}\fi\chkspace}
\def\Dpl{\relax\ifmmode {\rm D}^+ \else {D$^+$}\fi\chkspace}
\def\D0{\relax\ifmmode {\rm D}^0 \else {D$^0$}\fi\chkspace}
\def\Kst{\relax\ifmmode {\rm K}^* \else {K$^*$}\fi\chkspace}
\def\K0{\relax\ifmmode {\rm K}^0_s \else {K$^0_s$}\fi\chkspace}
\def\Kpl{\relax\ifmmode {\rm K}^+ \else {K$^+$}\fi\chkspace}
\def\Kstz{\relax\ifmmode {\rm K}^{*0} \else {K$^{*0}$}\fi\chkspace}